# Perspectives on 6G Architectures

Rainer Liebhart, Mansoor Shafi (*Life Fellow*, *IEEE*), Harsh Tataria *(Member, IEEE)*, Gajan Shivanandan (*Member*, *IEEE*), and Devaki Chandramouli (*Senior Member*, *IEEE*)

*Abstract*—Mobile communications have been undergoing a generational change every ten years. While 5G network deployments are maturing, significant efforts are being made to standardize 6G, which is expected to be commercially introduced by 2030. This paper provides unique perspectives on the 6G network (radio and core) architecture(s) from the anticipated 6G use cases to meet the necessary performance requirements. To cater for the key 6G use cases, the 6G architecture must integrate different network-level functions in a multiplicity of virtual cloud environments, leveraging the advancements of distributed processing, artificial intelligence, and securely integrating different sub-networks e.g., terrestrial, and non-terrestrial networks into the overall 6G network. This paper characterizes the impact of 6G architectures from a deployment perspective with backwards compatibility in mind.

## I. INTRODUCTION

Wireless communication is an integral part of our daily lives and the social fabric. Applications connecting humans to machines, and machines to machines are fuelling explosive growth in key performance metrics, e.g., data rates, latency and reliability. In response, mobile systems have undergone a generational change almost every 10 years. Whilst 5G deployments are maturing globally and 5G-Advanced (3GPP Release 19) is being standardized, considerable efforts are underway to research and eventually standardize 6G (a.k.a. IMT-2030) systems that are likely to be introduced from 2030 onwards. To this end, many research and industry white papers have appeared describing the 6G vision, performance requirements, enabling technologies, challenges, and architectural principles [1-6]. The ITU-R has approved a new Recommendation on the framework of 6G systems [7] World Radiocommunication Conference (WRC)-2023 has identified new candidate bands 4 500 – 4 800 MHz, 7 125 – 8 400 MHz and 14.8 – 15.35 GHz to be considered at WRC-27. However, 6G will also operate in existing bands ranging from sub-1 GHz, mid bands (3-8 GHz) and mmWave bands 24-100 GHz.

The 6G system needs to support a wide range of use cases and will thus need to support diverse and integrated network topologies, including terrestrial and non-terrestrial systems (e.g., fixed and mobile satellite systems), unmanned aircraft vehicle systems, autonomous driving systems and different categories of machine-type communications with integrated sensing capabilities. Sensing can be integrated in 6G on various levels, for example by using the same radio sites, sharing same frequency bands or re-using same hardware (this is not further elaborated in the paper). The sensing location accuracy would be dependent on the carrier bandwidth and radio access network (RAN) densification. The 6G architecture targeting the 6G services will have a flexible core network architecture that integrates different network functions in optimized virtual cloud environments, maximizing the benefits of artificial intelligence while securely operating the integrated terrestrial and non-terrestrial networks.

Unlike the existing literature, this paper provides unique concepts and perspectives on both the radio and core networks of 6G and characterizes the impact of these concepts and perspectives on the overall 6G system from a deployment perspective. While presenting the prospects of a 6G architecture, novel concepts on service-based architecture in RAN and core networks, application-aware Quality of Service (QoS), applications of AI and ML principles, the concept of network of networks, layer 2 and 3 protocol designs, RAN densification and security are discussed, weaving concepts from theory to reality.

## II. OVERVIEW OF 6G SERVICES AND KPIS

This section gives an overview of potential 6G services which will have a defining impact on the technologies and architecture for 6G. The use cases for IMT 2030 are categorized as [2,7]:

**Immersive media**: Extended reality (XR) encompasses VR (Virtual reality), AR (Artificial reality) and MR (Mixed Reality) and is of increasing interest in entertainment, medicine, education, and manufacturing. Truly immersive AR, XR media streaming, 16K UHD video require about 0.44 Gbps throughput [4].

**Digital twins**: This allows 6G to be used to replicate the physical world in a digital world in real time. This is applicable for the management of smart cities, where the digital twin applications may be used to manage transportation, utilities, forecast the impact of various events etc.

**Integrated sensing and communication**: This could be used for imaging, mapping, precise object detection, recognition, and estimation, etc.

**Contiguous and ubiquitous connectivity** that in turn will drive many use cases by addressing the challenge of connectivity, coverage, and capacity such as an integration between non terrestrial and terrestrial networks.

**High fidelity holographic society:** Advances in high resolution rendering and wearable technologies could make holographic telepresence the mode of choice for future communication, including multi-model communication for teleoperation, haptic and the tactile Internet. Data rates for holograms are more than 1 Tbps ([2], [4]) that are best available in sub-THz bands, which are not in the list of bands to be considered at WRC-27.

**Enhanced machine type communication:** One important use case for 6G is machine to machine communication. The performance targets for this kind of communication well exceed the current targets for human to human or human to machine communication, requiring e.g., a huge connection density.

**Enhanced broadband:** Wireless access points in metro stations, shopping malls, and other public places will be deployed as part of ultra-dense radio networks and providing extreme high data rates, low latency, and high positioning accuracy.

Many papers from industry and academia have provided views on the KPI metrics for 6G such as [1-4,6,7,9]; these will not be reviewed here. However, the metrics for 6G performance are truly formidable and have an impact on the technologies and architectures to be used for 6G networks – the main theme of this paper.

## III. MOTIVATION, DRIVERS, AND CONCEPTS FOR A 6G ARCHITECTURE

The use cases of 6G and their KPIs may not be efficiently provided by the 5G System (see [3],[4],[9]). The 6G use cases will require a true convergence of communication and computing, enabling the user's device to use computing power in the network efficiently. The 6G network will support a wide variety of sub-networks consisting of fixed base stations, mobile base stations, non-terrestrial base stations at various altitudes, different air mobility systems etc. These types of sub-networks are an integral part of the 6G architecture, allowing for seamless mobility between different coverage layers. We anticipate following main principles a future 6G network will follow:
- Adopting service-based architecture (SBA, see [10]) principles to the RAN
- Enabling application aware QoS
- Integration of improved and new sustainability metrics
- Extended use of Artificial Intelligence (AI) and Machine Learning (ML)
- Introducing the concept of "Network of Networks" including a densified RAN
- RAN densification via distributed massive MIMO
- Adopting new security principles

In the following we provide a closer look into these key technology areas. The main drivers, and potential concepts for a 6G architecture are followed by a high-level overview of the 6G architecture.

## IV. PROSPECTS OF A 6G ARCHITECTURE

At this stage it is premature to provide a functional architecture of a future 6GC as in 5G [10], our initial view is reflected in Figure 4(c).

This is because basic decisions have not been made so far: i.e. the functional split inside the 6G RAN inside the 6G Core Network (CN) and between RAN and CNs. Exploring the need to adopt concepts such as SBA to the RAN and to the RAN-CN interface might be natural. However, the functional and protocol details and their potential consequences need to be carefully studied before making any decisions and moving into the standardization phase. Nevertheless, some principles a 6G system must be based on seem to be clear:
- Smooth migration to 6G and Interworking with 5G.
- Supporting fully virtualised and non-virtualised network functions and components.
- Supporting applications requiring integration of heterogenous cloud environments: Far Edge, Edge, Regional and Central Clouds (Figure 2).
- Supporting adaptive QoS, best effort QoS and QoS aware applications requiring Gbps throughput rates and extreme low latency for real time applications.
- Integration of heterogenous (sub-)networks.
- Distributed Massive MIMO
- Advancements in the integration of computing, routing, and storage capabilities.
- Enhanced means of exposing data and services to third parties.
- Leveraging "as-a-Service (aaS)" concepts such as Network as a Service, Platform as a Service, and Infrastructure as a Service.
- Dual steering, NTN, and TN integration.
- Enhanced Security principles to address 6G network and service requirements.

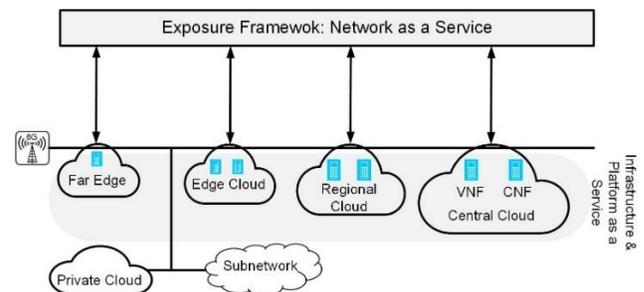

**Figure 1:** Heterogenous cloud deployments in 6G for Virtual Network Functions (VNF) and Cloud Native Functions (CNF)

### A. Cloud-Native Architecture for Radio Access Network and Core Network

Mobile networks are defined by functionalities divided into the CN and the RAN which are connected through open, interfaces. A key architectural change in 5GC was the transition to a cloud native service-based approach [10]. This trend will further extend towards the edge and the radio access in 6G with the benefit of enabling end-to-end deployments using a harmonized, cloud-based framework with common operational tools. The 6G network needs to accommodate a diverse set of use cases, services, and access technologies with different topologies. Cloud-based service delivery platforms will be diversified into; on premises, edge, core and public clouds through different processing and service capabilities matching the needs of these services. The target architecture for 6G should be flexible enough so that network functions and services can be deployed across different cloud platforms. Edge sites could host time sensitive radio related, e.g., Distributed Unit (DU), processing functions that benefit from HW acceleration. However, the flexibility of locating lower layer radio processing is limited by delay and capacity constraints, creating high demands on the transport network. The regional cloud could host less time sensitive functions e.g., the Central Unit (CU) implementing higher layers of the radio protocol stack and core functions like distributed UPF or SMF. As a consequence of the transition to a cloud-native service-based approach in the CN and in future in RAN, different functional splits between CN and RAN can be investigated in 6G. This could lead to an architecture where different functions that are traditionally located in CN and RAN domains are collocated in one cloud or even one network function if this is seen feasible (Figure 2). This will allow for higher deployment flexibility, e.g., flexible placement of signalling and user plane functions close or far away from the radio site depending on economic needs and application demands.

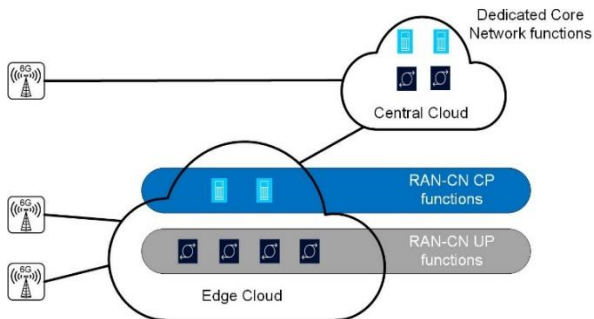

**Figure 2:** RAN-Core Control Plane (CP) and User Plane (UP) functions

### B. Application Aware QoS

A tighter coordination between applications and communication systems must be leveraged to improve end user experience. Real time immersive services are quite sensitive to packet loss, data rate and delay fluctuations. The need to support extreme throughput, real-time services such as holographic and Mixed-Reality (MR) communications, coupled with stricter security requirements, requires mobile networks to be aware of the QoS needs of applications. Low latency applications may also suffer from latency increase due to congestion that can occur either because of sudden link capacity changes or greedy traffic flows sharing the same link. Although real-time applications can adapt the transmission rate to sustain Quality of Experience (QoE), they are driving "blind". The current concept of guaranteed bit rates can run into scalability issues as it consumes lots of resources to sustain high data rates and low latency. To improve scalability, adaptive bit rate mechanisms and application aware QoS should be considered for 6G to effectively use resources. This will enable RAN to adapt the bit rate within the given range while enabling improved capacity, optimal end user experience and also reduced signalling between RAN and CN.

### C. Sustainability

Sustainability is a major requirement of the 6G architecture [7]. Thus, sustainability and energy efficiency are goals that need to be looked at from a holistic point of view for 6G; here the overall objective is to design an end-to-end architecture and protocols enabling an optimized, scalable and energy efficient operation of the whole network.

### D. Application of AI and ML Principles

AI/ML technologies are the essential ingredients for the 6G architecture. These technologies will help unlock the full potential of 6G systems, empowering them with new network automation capabilities, boosted performance and enhanced energy efficiency. Additionally, new services and applications employing AI/ML will impose new requirements for the 6G architecture.

There are various other emerging use cases that may require resource-constrained devices and are based on data-driven decisions, which will require new services such as compute-as-a-service (CaaS), general-purpose AI-as-a-service (AIaaS), and AI-assisted vehicle-to-everything (V2X). Each of these services is, in their turn, enabled by key technologies that are building blocks of the future 6G architecture.

### E. Network of Networks

The 6G network will comprise of many heterogeneous, multi-domain sub-networks serving different purposes and use cases. These networks are connected to a single 6G Core. This concept is called a "Network of Networks" (NoN), see Figure 3 and [3].

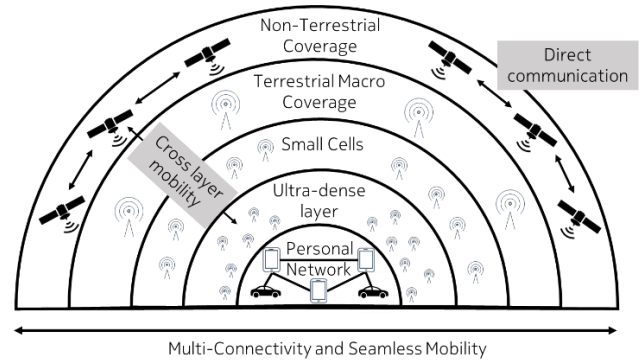

**Figure 3:** Network of Networks

Sub-networks in this context can be e.g., wireline or wireless networks using different types of technologies, e.g., 3GPP cellular and IEEE Wi-Fi. In future it will not be sufficient to allow all these different types of sub-networks to simply connect to a 6G Core or to each other, but to combine diverse access and deployment options to aggregate capacity, increase reliability, extend security, and allow for a seamless connectivity and interworking experience wherever and whenever a device connects via one or even multiple access networks simultaneously. This can be enabled by automatic or manual selection of the best fitting sub-network fulfilling certain requirements from applications, machines, or humans at a certain time and at a certain location. Intelligent network and access type selection can consider application needs, location, date and time, communication history and self-learning methods based on AI/ML. The devices can be connected to several of these sub-networks simultaneously and choose (with the help of the network or application layer) the one that fits best with respect to the particular demands of the device, human being, or applications. With the concept of NoN 6G promises to achieve such kind of integration, which will make the 6G system look like one unique big network, although consisting of potentially many diverse sub-networks of different sizes and capabilities.

### F. RAN densification via distributed Massive MIMO

A densified RAN is a key ingredient to address the requirements of 6G use cases envisaging ubiquitous connectivity and extremely high data rates with a finite amount of spectrum resource and energy consumption. Distributed massive MIMO (a.k.a. "cell-free" massive MIMO) provides a way to implement RAN densification by the introduction of coordinated distributed access points over a wide area, coherently serving many users equipment. This provides a higher instantaneous signal-to-interference-plus-noise ratio at a given user equipment and yields higher diversity gain providing coverage homogeneity.

By now, a vast literature on distributed massive MIMO systems exists, see [5,11] and references therein. However, most of the work do not present solutions that are scalable to real world implementation and deployment considerations, which seems to be the main bottleneck of RAN densification with distributed antennas. In what follows, we describe the three critical challenges:

1. To facilitate coordination between different access points, a centralized processing unit (CPU) is necessary from/to which signal paths are physically routed via fibre. This brings a significant burden to the scalability of the system while increasing the cost, which varies between different market conditions. The uptake of approaches to alleviate the cost of distributed cooperating antennas has been slow with directional transmission (as for fixed wireless access) in point-to-point links and integrated access/backhaul proving as meaningful trialled alternatives, naturally with a new set of challenges.
2. Each CPU naturally has a fixed-level of computation capacity and can host up to a finite maximum number of cluster processors, where a cluster processor can be a software-defined virtual network function running on the optimized hardware that implements layer 1 signalling for the corresponding user equipment, e.g., channel estimation, interference suppression and payload data decoding in the uplink, as well as computation of the downlink beamformers to be transmitted *jointly* by the cluster of access points. Front-haul signalling overheads can be reduced considerably by the design of optimal routing of the paths from the access point to the CPUs, as well as by optimizing the placement of cluster processors in the CPU. In the CPU then, for a given network topology, the load balancing and the allocation of computation resources must be jointly optimized. For instance, if the cluster processor for a certain user is allocated to a geographically remote CPU, the data between the CPU and the access points forming the cluster of that specific user must travel across many hops in the fronthaul network, thus generating a higher overall load.
3. Different types of synchronization are required for distributed access points to work coherently: A common frequency reference is necessary, and if joint reciprocity-based beamforming is to be used on downlink, access points also need to be phase aligned. As a first option one can distribute the RF signal itself from a CPU over fibre, such that no oscillator circuits and associated logic are needed at the access point. On the other hand, only distributing the baseband data, allows for a low-frequency reference carrier that drives local phase-locked loop oscillators at the access points. Considering these, joint reciprocity calibration can be performed over-the-air.

A distributed massive MIMO-enabled RAN design can assist in delivering efficient integrated sensing and communications. (ISAC). Sensing capability aims to provide localization, recognition and communications. In this regard, distributed massive MIMO-enabled RAN can assist sensing with two-levels of design methodologies:

- *Frame-level ISAC*: Sensing supported by default/pre-defined frame structures, e.g., for 5G-NR.
- *Network-level ISAC:* Distributed/networked sensing supported by state-of-the-art wireless architectures, e.g., Cloud RAN (C-RAN) as in 5G NR.

A detailed breakdown of challenges and solutions for both levels of design methodologies are provided in [12].

### G. 6G Security Principles

Security is an intrinsic part of the future 6G architecture [8,13]. The potential 6G threat vector has many more components relative to 5G. Large numbers of sensors, new human to machine interfaces build opportunities of threat. Additionally, the various sub-networks that may not have the same robustness against attacks (by allowing malicious identifications), as the main core network, disaggregated architectures, open interfaces all allow for the possibility of increased attacks. Below we describe some security enablers for 6G.

*Service Based Architecture Security*

Advances in 5G security took a major step forward with the implementation of the Service Based architecture (SBA) supporting secure and authorised communication for signalling between network functions (NFs) (e.g. mutual Transport Layer Security (mTLS) and Open Authorisation (OAUTH2.0)). This can be expected to carry forward to ensure interoperability as networks upgrade to support 6G.

*Non-Access Stratum (NAS) Security*

New concepts like modular NAS as described in the section on 6G Architecture and Migration further below enable independent security (integrity protection and encryption) for different NAS modules terminating in different Network Functions. As an example, the 6G Mobility Management function can use different length of security keys or different security algorithms for the NAS MM module than a 6G Session Management function for the NAS Session Management (SM) module. These functions can be part of the same security domain, e.g., in one network, or of different security domains, e.g., one in the home network and the other in the visited network or in a private network, also supporting a zero-trust architecture.

*Infrastructure and Platform Security*

With Cloud Native deployments being extended from 5G to 6G supporting infrastructure – security for software only deployments and underlying Containers as a Service (CaaS) environments need to be addressed. A range of techniques can be applied here - end to end code validation across supply chain for NFs, Infrastructure scanning (i.e., malware detection), and Physical infrastructure environments (i.e., Trusted Platform Module implementation). These techniques together with secure networking will help ensure security of the underlying transport and infrastructure supporting the 6G core network and its associated network functions.

*Vulnerability and Anomalous behaviour detection*

Security implementations will need to consider technologies that can be used to detect and manage unexpected security related events. As threats emerge specifically engineer with AI/ML techniques, so must the detection of such. As such a large portion of the 6G security is expected to be based on AI/ML techniques that can ingest a large variety of data including traffic (payload), performance stats and counters (time of day behaviours) and traffic flows. AI/ML will also automate closed-loop security operations combined with crowd sourced data from service endpoints through to eventual enforcement of policy (blocking). Dealing with analysis of the avalanche of traffic may need to be offloaded to specific network controllers embedded in the 6G network functions, rather than bespoke firewalls or other such enforcement points.

*Transactional Trust, Privacy and Security Everywhere*

Trust and Privacy are expected to be a key part of transactions with potentially the need for KPIs or other measures to quantify and measure trust and privacy based on use-case. The use of Distributed Ledger Technology (DLT) and blockchains to provide access validation and leverage blockchains to ensure reliability across infrastructure. Pervasive security in the face of connected things – to ensure security for these devices. Devices will seek to offload tasks and become more power efficient, this includes securing power saving techniques such as those envisaged for energy harvesting devices.

*Post Quantum Cryptography*

Traditional Public Key Cryptography, also known as Asymmetric Cryptography, is widely used in mobile networks and has been shown to be vulnerable to Shor's algorithm running on a Quantum Computer of sufficient capacity, referred to as a Cryptographic Relevant Quantum Computer (CRQC). NIST has standardized PQC (Post Quantum Cryptographic) asymmetric algorithms [14] which are believed to be resistant to attacks from both classical computers and CRQCs. Work is currently ongoing in IETF to adopt these NIST PQC algorithms (FIPS-203 for Key Encapsulation, FIPS204&205 for Digital Signatures) into security protocols and certificates (e.g., TLS, IPSec, X.509 etc.) and once completed profiles of these IETF protocols/certificates are expected to be adopted in the first release of 6G mobile networks.

*H. Dual Steering for Integration between NTN and TN*

Ubiquitous connectivity requires access to NTNs. Here a device camps simultaneously on NTN and TN as shown in Fig 4(a) and the traffic can be steered either via NTN or TN access. The dual steer device is registered with a 6G mobility management function, via different 3GPP accesses, requiring multiple access attachments. The user plane sessions for both UEs are anchored at the same UPF to enable steering, switching, and splitting of traffic via the NTN and TN access. This implies that the sessions are pre-established via both NTN and TN access, UPF steers the received downlink traffic either via TN or NTN. In case of mobility between TN and NTN, service interruption is minimized as the session is already established in the target domain and the traffic can be automatically transmitted and received via the target access network. This also enables seamless integration and service continuity between TN and NTN. Steering and switching is managed by the UPF for the downlink and by the dual steer layer in the device for the uplink thus it is also transparent from the radio network perspective unlike radio-based aggregation solutions.

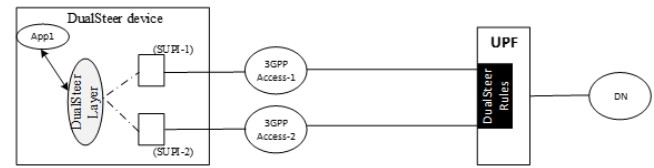

**Figure 4(a)**: Dual Steering between non-terrestrial network (NTN) and terrestrial network (TN)

*I. 6G Architecture and migration towards 6G*

A key learning from 4G to 5G migration was that many architecture options were specified leading to deployment fragmentation.

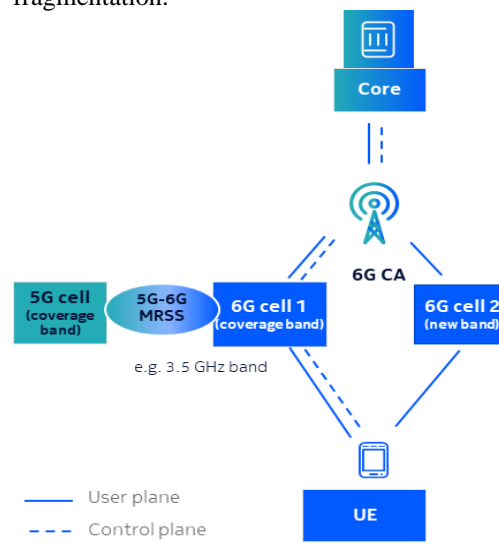

**Figure 4(b)**: 6G Standalone system

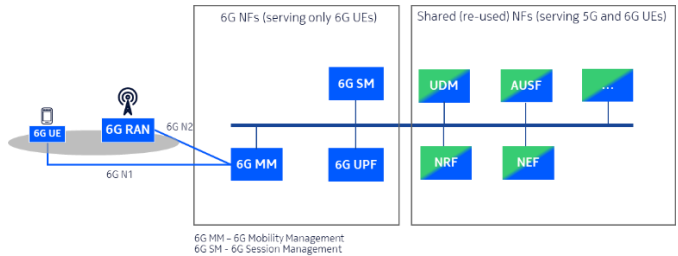

**Figure 4(c):** 6G system architecture

Therefore, with 5G to 6G migration, the goal is to have a *single* standalone system architecture specified. The 6G RAN is expected to be based on the *same* principles (e.g., Orthogonal Frequency Division Multiple Access (OFDMA) and Resource Block (RB) structure), as 5G-6G Multi-Radio Spectrum Sharing (MRSS) may be applied to allow for spectrum sharing with 5G bands. This is expected to be more efficient compared to Dynamic Spectrum Sharing (DSS) used for 4G-5G spectrum sharing. MRSS performance will not be limited by interference and overheads imposed by LTE cell reference signals (CRS) in LTE and 5G-6G MRSS can leverage the forward compatible 5G physical layer flexibility for efficient sharing. 6G may leverage different numerologies, support increased bandwidth utilization and support higher number of HARQ processes to allow for higher fronthaul latency enabling greater flexibility in 6G RAN function placement. Figure 4(b) above depicts the preferred target architecture for 6G migration using an example of a MRSS for a 3.5 GHz radio carrier supporting 5G and 6G, and Carrier Aggregation (CA) with a 6G carrier.

6G RAN is deployed standalone in conjunction with the CN that may comprise of 5G NFs shared between 5G and 6G and new NFs introduced for 6G as in Figure 4(c). 6G Mobility Management (MM) is expected to be a new NF due to foreseen 6G radio specific changes, leading to a 6G NAS protocol being terminated in the 6G MM. Also, 6G SM is expected to be a new NF due to the need of a new 6G NAS protocol, e.g., supporting the adaptive QoS framework as explained in this paper. Furthermore, support of enhanced modularity of NAS functions is a concept to be further investigated in 6G. This modular NAS concept with independent security terminations of the NAS modules in different NFs brings following benefits:

- Allows independent evolution of Mobility Management, Session Management, and other NAS modules in the future.
- Enables independent security for different NAS modules. As an example, 6G MM can use different length of security keys than a 6G SM.
- When the NFs are in different domains, e.g., VPLMN, HPLMN, MVNO (Mobile Virtual Network Operator), message content can be transmitted securely between the respective NF and the UE thus also enabling zero trust architecture.

Some NFs like NRF (Network Repository Function), NEF (Network Exposure Function) are expected to be "G" agnostic and can be re-used with necessary updates.

As in Figure 4 (b), 6G intra-RAT (CA) and/or 6G-6G dual connectivity (DC) can be used to combine capacity and coverage bands from bands newly allocated to 6G or bands that are deployed in 5G and leveraged by 6G via MRSS.

In an initial phase of 6G, CA may be the preferred choice. DC may be used for aggregating 6G carriers at different sites. CA support is expected to be simpler from the UE perspective as it does not necessarily always require dual UL support from UE to work and timings of two cells aggregated through CA are typically very close to each other. The SA architecture allows the RAN control plane to be based on a new 6G RRC, fully optimized for the new 6G RAN.

In addition to migration from 5G to 6G, also interworking between 6G and earlier technologies like 5G is quintessential as it is expected that 6G coverage is sparse initially compared to the existing 5G deployments. In order to support 5G-6G interworking with seamless service continuity, when the device moves between 5G and 6G within an operator's network, device must be capable of supporting single registration-based interworking and the network must support handover between 5G and 6G radio and a common IP anchor.

## V. CONCLUSION

This paper described new use cases and applications anticipated for the upcoming 6G era. Some of these applications are extended reality, enhanced machine type communication, and extreme broadband. The paper explained main pillars of a future 6G architecture which are in our view: adopting cloud native architecture principles to the Radio Access Network, enabling application aware QoS, integration of new sustainability and security measures, extended use of AI/ML and introducing new concepts like NoNs. Migration and interworking aspects are discussed to explain how the new 6G system can be introduced in a smooth manner using techniques like highly optimized multi-radio spectrum sharing between 5G and 6G.

## REFERENCES

[1] H. Tataria, et al, "6G Wireless Systems: Vision, Requirements, Challenges, Insights, and Opportunities," in Proceedings of the IEEE, vol. 109, no. 7, pp. 1166-1199, July 2021.
[2] H. Tataria, et al, "Six Critical Challenges for 6G Wireless Systems: A Summary and Some Solutions," in IEEE Vehicular Technology Magazine, vol. 17, no. 1, pp. 16-26, March 2022.
[3] "Expanded 6G vision, use cases and societal values", D 1.2, HexaX, Apr 2022; https://hexa-x.eu/deliverables/.
[4] '6G, the next hyper connected experience for all', Samsung, Dec 2020; https://news.samsung.com/global/samsung-unveils-6G-spectrum-white-paper-and-6G-research-findings.
[5] H. Q. Ngo, A. Ashikhmin, H. Yang, E.G. Larsson, T.L. Marzetta, "Cell-free massive MIMO versus small cells," IEEE Transactions on Wireless Communications, vol. 16, no. 3, pp. 1834-1850, Jan. 2019.
[6] "6G Technologies", Next G Alliance: An ATIS initiative, June 2022; https://nextgalliance.org/white_papers/6g-technologies/.
[7] ITU R M 2136, Framework and overall objectives of the future development of IMT for 2030 and beyond. Nov. 2023.
[8] V. Ziegler, et al: "Security and Trust in the 6G Era" in IEEE Access, vol. 9, pp. 142314-142327, October 2021.

[9] "Roadmap to 6G", Next G Alliance: An ATIS initiative, February 2022; https://roadmap.nextgalliance.org/.
[10] System architecture for the 5G System (5GS); 3GPP TS 23.501, June 2022.
[11] H. Q. Ngo, G. Interdonato, E.G. Larsson, G. Caire, J. G. Andrews, "Ultra-Dense Cell-Free Massive MIMO for 6G: Technical Overview and Open Questions," arXiv preprint arXiv:2401.03898, Jan. 2024
[12] Liu, et al., "Integrated sensing and communications: Towards dual-functional wireless networks for 6G and beyond," IEEE J. Sel. Areas in Commun., vol. 40, no. 6, pp. 1728 – 1767, Mar. 2022.
[13] P. Porambage, D. Pamela Moya Osorio, G. Gur, M. Liyange, "The Roadmap to 6G Security and Privacy", IEEE Open Journal of the Communications Society, May 2021.
[14] 3PQC Standards FIPS 203/204/205, retrieved from https://csrc.nist.gov/publication/fips., August 2024.

**Rainer Liebhart** is a research project manager in Nokia and delegate in 3GPP SA and SA2. Rainer is (co-)author of around 130 patents in the telecommunication area, of several IEEE papers and co-editor of books on 4G and 5G. Rainer has an MS degree in Mathematics and over 30 years of industry experience.

**Mansoor Shafi** is a Fellow at Spark New Zealand, Wellington, 6011, New Zealand. He has published widely in cellular communications and contributed to wireless standards in the ITU and 3GPP. He is an adjunct professor at Victoria and Canterbury universities, New Zealand. He is a winner of multiple IEEE awards.

**Harsh Tataria** is with Tataria Consulting. He has held several positions in academia and radio communications industry in the UK, US, Sweden and New Zealand.

**Gajan Shivanandan** has over 20 years of experience in the cellular mobile industry across architecture and engineering roles. His current role is as End-to-end Architect (Mobile) for Spark NZ in Wellington. He has recently engaged with 3GPP standardization groups specifically 3GPP SA2 and RAN4. He holds a B.Tech in Information Engineering from Massey University in Palmerston North, New Zealand.

**Devaki Chandramouli** is a Nokia Bell Labs Fellow and Head of North American Standardization at Nokia. She serves as a Steering Group co-chair in the Next G Alliance and a delegate in 3GPP SA2. She has co-authored IEEE papers, has over 200 patents and authored books on 5G and 4G. Devaki holds a B.E and M.S degrees both in Computer Science.